\documentclass[onecolumn,11pt]{IEEEtran}

\usepackage{graphicx,epsfig,float,subfigure,epstopdf,xcolor}      
\usepackage[labelfont=bf]{caption}
\usepackage{amsfonts,amsmath,amssymb} 
\usepackage{booktabs} 
\usepackage{chemformula}
\usepackage{soul} 

\newtheorem{lemma}{Lemma}
\newtheorem{proposition}{Proposition}
\newtheorem{corollary}{Corollary}
\newtheorem{fact}{Fact}

\newtheorem{remark}{Remark}

\newtheorem{assumption}{Assumption}

\def\begcen{\begin{center}}
\def\endcen{\end{center}}

\def\L2{{\cal L}_2}
\def\L2e{{\cal L}_{2e}}

\def\x{{x}}

\def\begequarr{\begin{eqnarray}}
\def\endequarr{\end{eqnarray}}
\def\begequarrs{\begin{eqnarray*}}
\def\endequarrs{\end{eqnarray*}}
\def\begarr{\begin{array}}
\def\endarr{\end{array}}
\def\begequ{\begin{equation}}
\def\endequ{\end{equation}}

\def\begdes{\begin{description}}
\def\enddes{\end{description}}
\def\begenu{\begin{enumerate}}
\def\begite{\begin{itemize}}
\def\endite{\end{itemize}}
\def\endenu{\end{enumerate}}

\def\lef[{\left[\begin{array}}
\def\rig]{\end{array}\right]}

\def\begcen{\begin{center}}
\def\endcen{\end{center}}
\def\begrem{\begin{remark}\rm}
\def\endrem{\end{remark}}
\def\begassum{\begin{assumption}}
\def\endassum{\end{assumption}}
\def\begassums{\begin{assumption*}}
\def\endassums{\end{assumption*}}
\def\begassu{\begin{ass}}
\def\endassu{\end{ass}}
\def\beglem{\begin{lemma}}
\def\endlem{\end{lemma}}
\def\begcor{\begin{corollary}}
\def\endcor{\end{corollary}}
\def\begfac{\begin{fact}}
\def\endfac{\end{fact}}

\def\L2e{{\cal L}_{2e}}

\def\begsubequ{\begin{subequations}}
\def\endsubequ{\end{subequations}}

\usepackage{xcolor}

\graphicspath{{}{img/}{Figures/}{Bios/}}
\begin{document}
\title{Exploiting Monotonicity to Design an Adaptive PI Passivity-Based Controller for a Fuel-Cell System}
\author{Carlo~A.~Beltr\'an,~\IEEEmembership{Member,~IEEE,}
	Rafael~Cisneros$^\star$,~\IEEEmembership{Member,~IEEE,}
	Diego~Langarica-C\'ordoba,~\IEEEmembership{Senior~Member,~IEEE,}
	Romeo~Ortega,~\IEEEmembership{Life~Fellow,~IEEE}~and~
	Luis H. D\'iaz-Saldierna,~\IEEEmembership{Member,~IEEE}	
    \thanks{$^\star$Corresponding author.}
	\thanks{C.A. Beltr\'an and D. Langarica-C\'ordoba are with  Faculty of Sciences, UASLP, 78295 San Luis Potos\'i, Mexico. E-mails: \ a332951@alumnos.uaslp.mx and diego.langarica@uaslp.mx.}
	\thanks{R. Cisneros and R. Ortega are with 
		Departament of Electrical and Electronics Engineering, ITAM, 01080 Mexico City, M\'exico. Emails: \{rcisneros,\;romeo.ortega\}@itam.mx.}
		\thanks{L. H. Díaz-Saldierna is with Instituto Potosino de Investigaci\'on Cient\'ifica y Tecnol\'ogica (IPICyT), 78216 San Luis Potos\'i, M\'exico. E-mail:ldiaz@ipicyt.edu.mx.}
}


\maketitle

\begin{abstract}
We present a controller for a power electronic system composed of a fuel cell (FC) connected to a boost converter which feeds a resistive load. The controller aims to regulate the output voltage of the converter regardless of the uncertainty  of the load. Leveraging the monotonicity feature of the fuel cell polarization curve we prove that the nonlinear system can be controlled by means of a  passivity-based proportional-integral controller. We afterward extend the result to an adaptive version, allowing the controller to deal with parameter uncertainties, such as inductor parasitic resistance, load, and FC polarization curve parameters. This adaptive design is based on an indirect control approach with online parameter identification performed by a ``hybrid'' estimator which combines two techniques: the gradient-descent and immersion-and-invariance algorithms. The overall system is proved to be stable with the output voltage regulated to its reference. Experimental results validate our proposal under two real-life scenarios: pulsating load and output voltage reference changes.
\end{abstract}

\section{Introduction}
A clean and non-intermittent source that is helping in the goal of \ch{CO2} reduction is the hydrogen fuel cell (FC) \cite{Jiao2021}. This device efficiently converts chemical energy into electrical energy using an electrochemical reaction that consumes oxygen and hydrogen to generate electrical energy, water, and heat. Much study in recent years has focused on the Proton-Exchange Membrane FC (PEMFC), a type of FC that stands out for its high power density, rapid start-up, versatility, and low operating temperature, among others \cite{Wang2020}. Possible applications of PEMFCs are transportation electrification and microgrids 
\cite{Ogungbemi2021}. In addition, due to the non-linear relationship between the output current and voltage, it is required to employ a power converter as an interface between the PEMFC and a DC link, forming a FC system. This converter routes energy from the FC to the load at voltage levels compatible with the operation of the load. However, advanced control algorithms are necessary to drive the overall operation of this FC system, to ensure tight output voltage regulation despite load changes, to meet specific dynamic responses, and to have robustness against parameter uncertainties. 
%
Furthermore, each PEMFC exhibits a characteristic current and voltage relationship in a steady-state operation called polarization curve, which is a widely used diagnostic approach for evaluating the performance of a PEMFC \cite{AndrewDicks}. This plot can be mathematically characterized using empirical models that include nonlinear functions of the current and constant parameters to approximate the voltage drops experienced in a real PEMFC. For instance,  a nonlinear static model, which uses three parameters is detailed in \cite{Shahin2010}. Another case is the fifth-order polynomial model given in \cite{Hilairet2013}. Moreover, in \cite{pemfc2}, the well-known  Larminie-Dicks model is detailed, where the voltage losses (activation, ohmic, and concentration) are taken into account. Additionally, in \cite{YUZIECON2018}, a two-termed power function model is detailed, where the parameters are the open circuit voltage and two positive constants. In general, to accurately predict the operation of the FC, the knowledge of model parameters is crucial. A way to determine these values is through offline estimation with data-fitting procedures, performed before the system starts its operation. However, in a real setting, these parameters are sensible to several factors such as temperature, humidity, etc. As a result of that, they change slowly while the system operation is in progress. In this regard, online estimation provides a solution to deal with these variations by continuously updating these estimates while the system operates. 
For example, in \cite{Xing2019}, the parameters for a nonlinear PEMFC model are estimated online.
To obtain the regressor, its nonlinear parametric model is linearized using Taylor series expansion. 
Then, an adaptive estimator is designed to achieve exponential parameter convergence, proved via Lyapunov stability theory. Experimental results validate the correct performance of this estimator.
On the other hand, in \cite{Chaoui2021} an online parameter estimator of a PEMFC modeled by an equivalent electrical circuit is designed. Its objective is the assessment of a PEMFC remaining useful life. For this, a Lyapunov-based adaptive law is developed. It proves asymptotic stability and parameter convergence and its performance is validated through experimental results.








In this work, we consider a system composed of a PEMFC, a boost converter, and a load. As a control problem, the derivation of control strategies that permit the voltage regulation of the system poses a challenge due to the non-linearities describing the behavior of the FC. 
Besides, due to the non-minimum phase (NMP) behavior, the output voltage regulation of the boost converter is carried out indirectly, through current mode control (CMC) \cite{CMCFurukawa2022}. As widely reported, this issue is circumvented by a scheme consisting of two control loops, each one evolving in a different time scale: a ``voltage'' outer loop and a ``current'' inner loop.  The rationale of this control strategy is the following. The inner control loop regulates the current to a desired reference. On the other hand, the outer loop regulates the voltage to its setpoint by providing to the inner loop the corresponding reference of the current that makes possible such task.  
Traditionally, the voltage loop is implemented with a proportional-integral (PI) controller, and the current loop with linear or non-linear controllers, such as passivity-based control (PBC), backstepping, and sliding mode control, among others. In particular, for the system under consideration, the CMC scheme is employed in \cite{FCzuniga}, where the inner current loop is designed with backstepping control, and the outer voltage loop is designed using a classical PI action over the output voltage. Moreover, in \cite{FCmicro},  the current loop is designed with classical PBC, and the robustness of this CMC scheme is enhanced by online estimating of the parasitic resistance of the inductor and the load conductance using Immersion and Invariance (I\&I) \cite{bookastolfi}. In these previous works, Lyapunov stability is demonstrated, tight voltage regulation is obtained, and the PEMFC is modeled with a two-termed power function with their parameters estimated with offline data fitting procedures. On the other hand, a simplified scheme using a single control PI loop, based on passivity, is presented in \cite{pemfc2}, where offline estimation of the PEMFC parameters and online estimation of the load are performed to compute the equilibrium points required by the proposed controller. 
This approach exploits the monotonic nature of the polarization curve to design the control scheme. Relying on this property, the practical stability of the system operation is proven for the joint operation of the controller with the estimator. 



In this work, an improvement of the adaptive controller designed in \cite{pemfc2} is presented. This is done by adding the gradient-descent (online) algorithm \cite{booksastry}---a standard approach  in engineering applications---to estimate the FC parameters. The algorithm operates simultaneously with an I\&I algorithm, which estimates the converter parameters. This enables the online estimation of the equilibrium point required by the PI-PBC scheme. Besides, in this note, it is also proven exponential stability of the controller when all parameters are known and asymptotically stability of the overall adaptive system. Finally, we remark that as far as the authors' knowledge, there is no previous work in literature where the FC polarization curve is estimated online in closed-loop operation.

The rest of the paper is organized as follows. Section 2 presents the FC system under study and introduces the PI-PBC assuming the parameters are known. Then, in Section 3, this controller is turned adaptive with an online estimator based on I\&I and gradient-descent theory.  
Experimental results revealing the closed-loop performance of the system are presented in Section 4. Finally, in Section 5, the conclusions of the results are presented and suggestions for further research are provided.


\noindent \textbf{Notation.} $I_n$ corresponds to the $n\times n$ identity matrix. When clear from the context the arguments of the functions are omitted. Given a full-rank matrix $G(x)\in\mathbb{R}^{n\times m}$, with $n> m$, we denote its left annihilator as $G^\perp(x)\in\mathbb{R}^{(n-m)\times n}$ and its pseudo inverse as $G^+(x):= [G^\top(x) G(x)]^{-1}G^\top(x)$---i.e., they satisfy $G^\perp G=0$ and $G^+G=I_m$. We denote $v_i$ as a vector of the standard basis with its $i$-th element equal to one---its dimension can be inferred from the context. We refer to $\mathcal{L}_2$ as the set of square-integrable functions $f:\mathbb{R}_+\to\mathbb{R}$, namely, they satisfy $\int_0^\infty |f(t)|^2 dt<\infty.$ 

\section{System description and control}
\begin{figure}
	\centering
	\includegraphics{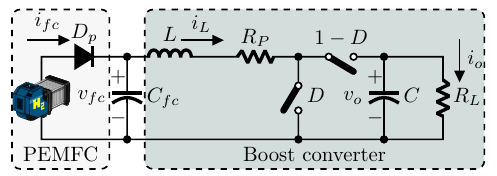}
	\caption{FC System under consideration.}
	\label{f1}
\end{figure}

The electrical circuit of the FC system under consideration is given in Fig. \ref{f1}. As can be seen from the figure, the system is composed of a PEMFC which feeds a load through a protective diode $D_p$ and a boost DC-DC converter. The converter regulates the output voltage to which the load is connected. This voltage is kept constant at a desired setpoint regardless of how much power it is being consumed by the load.

The model of the system in Fig. \ref{f1} is represented by the equations  \cite{pemfc2,pemfc1}
\begin{subequations}\label{pemfceq1}
	\begin{align}
		C_{fc}\dot v_{fc}= &i_{fc} - i_L, \label{pemfceq1a_1}\\
		L \dot i_L = & -R_p i_L+ v_{fc} - (1-D) v_o,\\
		C\dot v_o =& -v_o/R_L+ (1-D) i_L,
	\end{align}
\end{subequations}
where 	$C_{fc}$ and $C$ are the coupling fuel-cell capacitor and output converter capacitor, respectively. Also, $L$ is the converter inductance, $R_p$ is the inductor parasitic resistance, $R_L$ is the load resistance. The signal $D\in ( 0,1)$ corresponds to the converter duty cycle. On the other hand,  $v_o$ is the capacitor output voltage, whereas $v_{fc}$ and $i_{fc}$ are the fuel-cell voltage and current, respectively.  These two last variables relate to each other by means of the polarization curve \cite{YUZIECON2018}
\begin{equation}\label{pc}
	I_{fc}(v_{fc}):= i_{fc}=\left (\frac{E_{oc} - v_{fc}}{\theta_{s1}}\right)^{\frac{1}{\theta_{s2}}},
\end{equation}
where $E_{oc}\geq 0$ refers to the open-circuit voltage of the FC. Also, the parameters $\theta_{s1}$ and $\theta_{s2}$ are positive. 
According with the physical operation of the FC System, its current and voltage variables satisfy the following
 
\begin{itemize}
	\item[P1.] $v_{fc}$ is nonnegative and $E_{oc}-v_{fc}> 0.$
	\item[P2.]$i_{fc}$ and $v_o$ are nonnegative.
\end{itemize}

Therefore, P1 and P2 are standing assumptions throughout this note.

\begin{fact}\label{fct}
	The relation between the voltage and current  in the function $-I_{fc}$ is strongly monotonic. Namely, for any scalars $a$ and $b$  satisfying P1, there exists a constant $\alpha>0$ such that  the following inequality holds
	\begin{equation}\label{mon}
		(a-b)\Big[[-I_{fc}(a)]- [- I_{fc}(b)]\Big ]\geq\alpha (a-b)^2.
	\end{equation}
\end{fact}
\begin{IEEEproof}
	The derivative of $-I_{fc}$ with respect to $v_{fc}$ is
	$$\frac{d}{dv_{fc}}[-I_{fc}(v_{fc})]= \frac{1}{\theta_{s1}\theta_{s2}}\left(\frac{E_{oc}-v_{fc}}{\theta_{s1}}\right)^{\frac{1}{\theta_{s2}}-1},$$
	which is positive, proving the claim \cite{pavlov}.
\end{IEEEproof}

We now replace \eqref{pc} into \eqref{pemfceq1a_1}. The resulting equations  \eqref{pemfceq1} can be equivalently written as follows.
\begin{fact}
	The FC System \eqref{pemfceq1} can be represented by the dynamical system
	\begin{equation}\label{Phmodel}
		Q\dot x = [J_0 + J_1 u -R]x +  v_1 I_{fc}(x_1),
	\end{equation}
	where $u:=1-D$,
	\begin{align*}
		x&:=\begin{bmatrix}v_{fc}\\i_L\\v_o\end{bmatrix},\;Q:=\begin{bmatrix}C_{fc}&0&0\\0&L&0\\0&0&C\end{bmatrix},\;
		J_0:=\begin{bmatrix}0&-1&0\\1&0&0\\0&0&0\end{bmatrix},\\ J_1&:=\begin{bmatrix}0&0&0\\0&0&-1\\0&1&0\end{bmatrix},\;R:=\begin{bmatrix}0&0&0\\0&\theta_{r1}&0\\0&0&\theta_{r2}\end{bmatrix}.\;
	\end{align*}
 For convenience, we also define the following vectors of parameters $$\theta_{s}:=\begin{bmatrix}\theta_{s1}\\\theta_{s2} \end{bmatrix},\;\theta_{r}:=\begin{bmatrix}\theta_{r1}\\\theta_{r2}\end{bmatrix},\;\theta:=\begin{bmatrix}\theta_{r}\\\theta_{s}\end{bmatrix}.$$
\end{fact}

\subsection{Assignable equilibrium points}
The control objective is to regulate the converter output voltage $x_3$ to some reference $x_3^\star>0$.  According to that, the possible closed-loop equilibrium points are given in the following lemma.
\begin{lemma}\label{eq}
	The assignable equilibrium points of \eqref{Phmodel} and the associated constant input are those values in the set
	\begin{align}\label{eqp}
		\mathcal{E}&:=\left\{ (x,u) \in\mathbb{R}^4:x_1=E_{oc} - \theta_{s1}x_2^{\theta_{s2}},\;p(x_2,\theta)=0,\;x_3=x_3^\star,\; u=x_3^\star\frac{(\theta_{r2}-\theta_{r1})x_2 + x_1}{x_2^2+ (x_3^{\star })^2} \right\},
	\end{align} 
	where 
	$$p(x_2,\theta):= \theta_{r1}x_2^2 +\theta_{r2}(x_3^\star)^2 - x_2[E_{oc}-\theta_{s1}x_2^{\theta_{s2}}]. $$
\end{lemma}
\begin{IEEEproof}
	Notice that \eqref{Phmodel} can be written in the affine form $Q\dot x = f(x) + g(x)u$ with
	\begin{equation}\label{aff}
		f(x):= (J_0-R)x,\; g(x):=\begin{bmatrix} 0\\-x_3\\x_2\end{bmatrix}.
	\end{equation}
	A full-rank left-hand annihilator of $g(x)$ is
	$$g^\perp (x)= \begin{bmatrix}1&0&0\\x_1&x_2&x_3\end{bmatrix}.$$
	Therefore, at the equilibrium, the next relation holds
	\begin{equation}\label{equrel}
		g^\perp (x)f(x)=\begin{bmatrix}
		I_{fc} -x_2\\
		-\theta_{r1}x_2^2-\theta_{r2}x_3^2  + x_2x_1
		\end{bmatrix}=0.
	\end{equation}
From \eqref{pc}, we obtain the voltage $x_1$ as
\begin{equation}
    x_1=E_{oc}-\theta_{s1}I_{fc}^{\theta_{s2}}.
\end{equation}
Employing the first equation in \eqref{equrel} produces the equilibrium value of $x_1$ showed in $\eqref{eqp}$. Replacing it into the second equation of \eqref{equrel}, with $x_3=x_3^\star$, results in the second equation of $\mathcal{E}$.	
 The control input $u$ in \eqref{eqp} is obtained using the left inverse as follows  $$u=-g^+(x)f(x).$$
\end{IEEEproof}

\subsection{The full-information PI Passivity-based control}
The proposition below introduces the PI-PBC for \eqref{Phmodel}. As a first approach, we assume that all parameters in $\theta$ are known. An adaptive version of this controller is later presented, where the parameters are estimated online. For both the full-information PI-PBC and its adaptive extension, we assume the following.

\begin{assumption}\label{ass1}
	$x$ and $i_{fc}$ are measurable.
\end{assumption}
\begin{lemma}\label{pi}
	Consider the FC System modeled by \eqref{Phmodel}. Assume that all the parameters are known. Fix a desired, constant value for $x_3$ as $x_3^\star>0$ and compute from $\mathcal{E}$ the associated equilibrium vector $x^\star$. Consider the PI-PBC 
	\begin{subequations}\label{fipipbc}
		\begin{align}
			\dot{x}_c=&y_N (x),\\
			u=& -K_P y_N (x) - K_I x_c,
		\end{align}
	\end{subequations}
	where the input signal to the PI is defined as
	\begin{align}\label{yn}
		y_N(x)=& x_2^\star x_3- x_3^\star x_2.
	\end{align}
	For all positive constants $K_p$ and $K_I$ we have that all signals remain bounded ensuring the \textit{exponential} convergence
	$$\lim_{t\to \infty}\begin{bmatrix}
		x(t)\\x_c(t)
	\end{bmatrix} = \begin{bmatrix} x^\star\\x_c^\star\end{bmatrix},$$
	 where $x_c^\star =- K_I^{-1}u^\star$ with $u^\star\in\mathcal{E}$ the value of the control input at the equilibrium.
\end{lemma}

\begin{IEEEproof}
	We will first show that the system is stable. It follows modifying the proof of \cite[Prop. 2]{pimike} to include the presence of the term $I_{fc}$ and then invoke the monotonicity of Fact \ref{fct}.   
 
	From \eqref{Phmodel}, the error dynamics are
	\begin{align}
		Q\dot{ \tilde{x}} =& \big[ J_0 + J_1(\tilde u+u^\star ) -R  \big](x^\star + \tilde x) + v_1I_{fc} \pm v_1I_{fc}^\star,\nonumber\\
		 =&(J_0+J_1 u^\star-R) x^\star  + v_1I_{fc}^\star+(J_0+J_1 u-R)\tilde x +\tilde u J_1 x^\star  +v_1(I_{fc}-I_{fc}^\star),\nonumber\\ 
		=& (J_0+J_1u -R) \tilde x +J_1 x^\star\tilde u +v_1(I_{fc}-I_{fc}^\star),\label{tildx}
	\end{align}
	where $\tilde{()}:=()-()^\star$, $I_{fc}^\star:=I_{fc}(x_1^\star)$ and we use the equilibrium equation
	$$(J_0+J_1u^\star-R)x^\star + v_1I_{fc}^\star=0,$$
	to get the third identity. Now, we notice from \eqref{yn} that the passive output $y_N$ may be written as $$y_N=x^\top J_1 x^\star,$$ and moreover that $y_N(x^\star)=0$, hence	$$y_N(x)=y_N(\tilde x)= \tilde x^\top J_1 x^\star.$$
	
	Consider the Lyapunov function candidate for the system \eqref{fipipbc} and \eqref{tildx}
	$$V(\tilde x,\tilde x_c)= \frac{1}{2}\tilde x^\top Q\tilde x + \frac{K_I}{2}\tilde x^2_c.$$
	Its time derivative satisfies
	\begin{align*}
		\dot V=& -\tilde x^\top R \tilde x + \tilde uy_N (\tilde x) + K_I\tilde x_c y_N(\tilde x) + \tilde x_1[I_{fc}-I_{fc}^\star]. 
	\end{align*}
	From the inequality \eqref{mon} in Fact \ref{fct}, we have that 
	$$(x_1-x_1^\star) [ I_{fc}- I_{fc}^\star  ]=\tilde x_1 [I_{fc} - I_{fc}^\star]\leq -\alpha \tilde x_1^2.$$
	Therefore,
	\begin{align*}
		\dot V \leq & -\tilde x^\top R\tilde x-\alpha \tilde x_1^2 +\tilde u y_N(\tilde x) +K_I\tilde x_c y_N(\tilde x), \\
		=& -\tilde x^\top R\tilde x -\alpha \tilde x_1^2 + (u-u^\star)y_N(\tilde x) +K_I \tilde x_c y_N(\tilde x),\\
  =& -\tilde x^\top R\tilde x -\alpha \tilde x_1^2 + \big[-K_Py_N(x)-K_Ix_c +K_I x_c^\star \big] y_N(\tilde x) + K_I \tilde x_c y_N(\tilde x),\\
		=&-\tilde x^\top R\tilde x- \alpha \tilde x_1^2 - K_Py_N^2(\tilde x), \\
		\leq& -\tilde x^\top \mathrm{diag}(\alpha,\theta_{r1},\theta_{r2})\tilde x  - K_Py_N^2(\tilde x) , \\
		\leq& -\kappa |\tilde x|^2 - K_P (\tilde x^\top g_\star)^2,
	\end{align*}
	with $\kappa:=\mathrm{min}(\alpha,\theta_{r1},\theta_{r2})$ and, from \eqref{aff}, $g_\star:=g(x^\star)=J_1x^\star$. We conclude that $\tilde x$ and  $\tilde x_c$ are bounded. Consequently, $\tilde u$ and $u$ are also bounded.  Moreover, the error dynamics of the closed loop are 
 \begin{equation}
	\begin{aligned}
		Q\dot{\tilde x}=& [R- J_0u+J_1]x - g_\star [K_Pg_\star^\top\tilde x + K_I\tilde{x}_c]+v_1(I_{fc}-I_{fc}^\star),\\
   \dot {\tilde{x}}_c=& g_\star^\top \tilde x,
	\end{aligned}
 \end{equation}
	The second step of this proof consists in demonstrating exponential stability as in  \cite[proposition 2]{pidzonetti}. For, we consider the following function
	\begin{align*}
		W(\tilde x,\tilde x_c) =& V(\tilde x,\tilde x_c) + \epsilon K_I \tilde x^\top g_\star  \tilde x_c,\\
		=& \frac{1}{2}\chi^\top \begin{bmatrix} Q & \epsilon K_I g_\star \\ \epsilon K_I  g_\star^\top  & K_I\end{bmatrix} \chi, 		
	\end{align*}
	where $\chi:=\mathrm{col}(\tilde x,\tilde x_c)$  and a  constant $\epsilon>0$. The function $W$ is positive definite iff 
	\begin{align}\label{con1}
		Q-\epsilon^2K_Ig_\star g_\star^\top>0.
	\end{align}
	
	The time derivative of $W$ is 
	\begin{align*}
		\dot W =& \dot V + \epsilon K_I \tilde x^\top g_\star \dot{\tilde x}_c + \epsilon K_I \tilde x_c  g_\star^\top \dot{\tilde x}, \\
		\leq& -\kappa |\tilde x|^2 -K_P (\tilde x^\top g_\star)^2 + \epsilon K_I (\tilde x^\top g_\star)^2  +  \epsilon K_I \tilde x_c g_\star^\top \dot{\tilde x},\\
		=& -\kappa |\tilde x|^2 - (K_P -\epsilon K_I)(\tilde x^\top g_\star)^2+ \epsilon K_I\tilde x_c \big (g_\star^\top Q^{-1}[J_ou+J_1-R]  -K_P g_\star^\top Q^{-1}g_\star g_\star^\top \big )\tilde x    - \epsilon K_I^2  g_\star^\top  Q^{-1} g_\star  \tilde x_c^2,\\
		=& -\kappa |\tilde x|^2 - (K_P -\epsilon K_I)(\tilde x^\top g_\star)^2- \epsilon \tilde x_c s^\top \tilde x    - \epsilon K_I^2  g_\star^\top  Q^{-1} g_\star  \tilde x_c^2,
	\end{align*}
	where, to obtain the expression in the third line, we used the fact that the product $ g_\star^\top \dot{\tilde x}$ is
	\begin{align*}
		g_\star^\top \dot{\tilde x},
		=& g_\star^\top Q^{-1} \left\{ [J_0u+J_1-R]\tilde x - g_\star[K_p g_\star^\top \tilde x + K_I\tilde x_c ]  +v_1(I_{fc}-I_{fc}^\star) \right\},
	\end{align*}
 	with
	\begin{align}\label{cond}
	g_\star^\top Q^{-1} v_1 (I_{fc}-I_{fc}^\star)=0.
	\end{align}
	
	Also, to obtain the expression in the forth line, we defined
	$$s^\top:=-K_Ig_\star^\top Q^{-1}[J_0u+J_1-R]+ K_IK_p g_\star^\top Q^{-1}g_\star g_\star^\top, $$
	
	The inequality above can be written in the matrix form 
	\begin{align}\label{d_w}
		\dot W\leq - \chi^\top  M \chi,
	\end{align}
	where
	$$M:=\begin{bmatrix} \kappa I_3 + (K_P-\epsilon K_I)g_\star g_\star^\top & \frac{\epsilon}{2}s\\\frac{\epsilon}{2}s^\top &  \epsilon K_I^2 g_\star^\top Q^{-1}g_\star  \end{bmatrix}.$$
	The matrix $M$ is positive definite iff
	\begin{align}\label{con2}
		\kappa I_3  +& K_pg_\star g_\star^\top  -\epsilon  \left[ K_I g_\star g_\star^\top +\frac{1}{4K_I^2}s (  g_\star^\top Q^{-1}g_\star )^{-1}s^\top \right]>0. 
	\end{align}
	We conclude the proof noting that there exists a sufficiently small constant $\epsilon>0$ satisfying  \eqref{con1} and \eqref{con2}. Exponential convergence follows \cite[Theorem 4.10]{khalil}.
\end{IEEEproof}


\section{Main result}

\subsection{Estimation of $\theta$}
An estimator of $\theta$ is a dynamical system of the form
\begin{equation}\label{estim}
	\begin{aligned}
		\dot{\eta}(t)=& \chi_\eta(t,\eta(t),x(t)),\\
		\hat \theta(t)	=& \chi_{\theta} (t,\eta(t) ,x(t)),
	\end{aligned}
\end{equation}
\noindent where $\hat\theta$ is the estimate of $\theta$. Such that
\begin{align}\label{conv}
	\lim_{t\to\infty} \hat\theta(t) = \theta.
\end{align}
\noindent Here below, an estimation algorithm for $\theta$ in \eqref{Phmodel} is proposed. This combines two already reported estimation approaches: the I\&I technique \cite{bookastolfi} and the gradient-descent estimator \cite{booksastry} techniques. More precisely, the I\&I approach is employed to estimate $\theta_r$ whereas the gradient-descent estimator is implemented to identify $\theta_s$.

The next lemma introduces a linear regression equation (LRE) obtained from the polarization curve \eqref{pc}. This LRE is part of the estimator equations introduced below. The proof can be found in \cite[Lemma 4]{pemfc1}. 

\begin{lemma}
	Consider the algebraic relation in \eqref{pc} with $E_{oc}$ as a known parameter. Then, the next LRE holds
	\begin{align}\label{lre}
		Y(t)=\phi(t) \theta_{s2},
	\end{align}
	where
	\begin{equation*}
		\begin{aligned}
			Y&=\mathcal{F}\{\ln (E_{oc} - x_1)\},\\
			\phi&= \mathcal{F}\{\ln (i_{fc})\},
		\end{aligned}
	\end{equation*}
\end{lemma}
and the operator $\mathcal{F}\{\cdot\}$ is the stable, LTI filter 
$$\mathcal{F}:= \frac{\lambda p}{p+\lambda},\; \lambda>0.$$

Before introducing the estimation algorithm, the following excitation assumption is in order.

\begin{assumption}\label{ass2}
	$x_2,x_3$ and $\phi\not\in \mathcal{L}_2$.
\end{assumption}
\begin{proposition}[Parameter estimator]\label{prop1}
	 Let the estimator \eqref{estim} be composed of the following dynamics:
	\begin{itemize}
		\item[E1.] (\textit{Estimation of $\theta_s$}) The gradient-descent estimator:
		\begin{subequations}\label{gd_fc}
			\begin{align}
				\dot{\hat{\theta}}_{s2} =& \gamma \phi (Y  - \phi \hat\theta_{s2} ),\label{grad1}\\
				\hat{\theta}_{s1}=& (E_{oc} -x_1)i_{fc}^{-\hat{\theta}_{s2}},\label{grad2}
			\end{align}
		\end{subequations}
		for a positive gain $\gamma.$
		\item[E2.] (\textit{Estimation of $\theta_r$}) The I\&I estimator:
		
		\begin{subequations}\label{iandi_fc}
			\begin{align}
				\dot \xi_{1}=&-k_1x_2\left(-x_1  -\frac{1}{2}k_1Lx_2^3 + r_{1}x_2  + x_3 u\right),\label{estim1}\\
				\dot \xi_{2}=&-k_2x_3\left(-x_2 u -\frac{1}{2}k_2Cx_3^3 +r_{2}x_3 \right)\; \label{estim2},\\
				\hat\theta_{r1}=&-{k_1 \over 2} Lx^2_2+\xi_{1}\label{iandi_fc3},\\
				\hat\theta_{r2}=&-{k_2 \over 2} Cx^2_3+\xi_{2},
			\end{align}
		\end{subequations}
		where $k_1$ and $k_2$ are positive gains.
	\end{itemize}
	Fulfillment of Assumption \ref{ass2} ensures that the estimation error, defined as $e:=\hat{\theta}-\theta$, 
 is bounded  and
	\begin{align}
		\lim_{t\to\infty }e(t)=0,
	\end{align}
\end{proposition}

\begin{IEEEproof}
	The gradient-descent estimator in \eqref{grad1} is a standard estimation algorithm. Its error $e_{s2}$, defined as  $e_{s2}:=\hat{\theta}_{s2}-\theta_{s2}$, has the following dynamic
	$$\dot{e}_{s2}=-\gamma \phi^2 e_{s2}.$$
Assumption \ref{ass2} implies the convergence of $e_{r2}$. Moreover, using \eqref{pc} with $\hat \theta_{s2}$ instead of the actual parameter value $\theta_{s2}$, it is possible to obtain an estimate of $\theta_{s1}$ as in \eqref{grad2}.
	
	We now prove the convergence of the I\&I estimator. The estimation error  $e_{r1}$, defined as $e_{r1}:=\hat\theta_{r1} - \theta_{r1}$, has the following dynamics
	\begin{align*}
		\dot{e}_{r1}=& \dot r_{1} - k_1Lx_2\dot x_2,\\
		=&k_1x_2(x_1 - r_{1}x_2 +{k_1 \over 2}  L x^3_2  - x_3 u)- 	k_1x_2[-(\hat\theta_{r1}-\tilde\theta_{r1})x_2+x_1-x_3u],\\
		=&-k_1x_2(x_3 u  +\hat{\theta}_{r1}x_2 )- 	k_1x_2[-(\hat\theta_{r1}-\tilde\theta_{r1})x_2-x_3u],\\
		=&-k_1 x_2^2\tilde{\theta}_{r1},
	\end{align*}
	where \eqref{iandi_fc3} was used to obtain the third equality. With a similar procedure, the estimation error $e_{r2}:=\hat\theta_{r2} - \theta_{r2}$, evolves according to the following dynamics
	\begin{align*}
		\dot{{e}}_{r2}=& \dot r_{2} - k_2Cx_3\dot x_3,\\
		=&k_2x_3[x_2 u -  r_{2}x_3 +{k_2 \over 2} C x^3_3]- 	k_2x_3[-(\hat\theta_{r2}-\tilde\theta_{r2})x_3+x_2u],\\
		=&k_2x_3(x_2 u  -\hat{\theta}_{r2}x_3 )- 	k_2x_3[-(\hat\theta_{r2}-\tilde\theta_{r2})x_3+x_2u],\\
		=&-k_2 x_3^2\tilde{\theta}_{r2}.
	\end{align*}
	Again, since $x_2$ and $x_3\not\in\mathcal{L}_2$ by assumption, $e_{r1}$ and $e_{r2}$ converge to zero.
	
\end{IEEEproof}

\subsection{The proposed adaptive PI-PBC}
The full-information PI-PBC  of Lemma \ref{pi} depends on the equilibrium points. When parameters are available, the equilibrium are numerically computed by a root-finding procedure applied to the equilibrium equations of $\mathcal{E}$. On the other hand, when these parameters are unknown, a parameter estimation has to be performed. An estimate of the equilibrium point can be carried out by solving the equations of \eqref{eqp} that result from replacing the actual parameters with their estimate $\hat\theta$. In other words, for a given $\hat{\theta}$, the estimate of the equilibrium point, denoted as $\hat x^\star$, belongs to the set
\begin{align}\nonumber
	\hat{\mathcal{E}}:=&\left\{ x\in\mathbb{R}^3: x_1=E_{oc}-\hat{\theta}_{s1}x_2^{\hat{\theta}_{s2}}\right.,\\&\hspace{3cm}\;\left.p(x_2,\hat\theta)=0,~ x_3=x_3^\star \right\},\label{eseq}
\end{align} 
where the mapping $p(\cdot,\cdot)$  has been defined in  \eqref{eqp}. It can be noted that finding the elements in $\hat{\mathcal{E}}$ involves the search of a root of $p(x_2,\hat{\theta})=0$, which in principle, may not exist. In this sense, the Implicit Function theorem (see, for example, \cite[Section A.1]{bookisidori}) provides sufficient conditions for the existence of solutions of this equation.

\begin{lemma}\label{eqexistence}
	Let $x^\star\in\mathcal{E}$ be an assignable equilibrium point of \eqref{Phmodel} and assume that
	\begin{align}
		2{\theta}_{r1}x_2^\star + (\theta_{s2}+1)\theta_{s1}(x_2^\star)^{{\theta}_{s2}} -E_{oc} \not =0.  \label{C_2}
	\end{align}
	 Then, there exist   $\varepsilon>0$ and $\hat x_2^\star\in\mathbb{R}$ such that for any $\hat\theta$ satisfying $|\hat{\theta}-\theta|<\varepsilon$,  it holds that
	 $$p(\hat x_2^\star,\hat \theta)=0.$$
	 Namely, a solution of the second equation in \eqref{eseq} exists.
\end{lemma} 
\begin{IEEEproof}
The derivative of $p(x_2,\hat{\theta})$ with respect to $x_2$ is
\begin{align*}
	\frac{dp}{dx_2}(x_2,\hat{\theta})=&2\hat{\theta}_{r1}x_2 + (\hat{\theta}_{s2}+1)\hat{\theta}_{s1}x_2^{\hat{\theta}_{s2}} -E_{oc}.
\end{align*}
	We then evaluate the last expression in $(x_2,\hat{\theta})=(x_2^\star,\theta)$. The resulting equation is prevented from being zero by condition \eqref{C_2}. Consequently, the Implicit Function theorem guarantees the existence of a smooth function $h(\hat\theta)$, mapping each $\hat{\theta}$ in a neighborhood of $\theta$ to a point $\hat{x}_2^\star$ in a neighborhood of $x_2^\star$ such that
	$$p(h(\hat\theta),\hat{\theta})=0.$$
\end{IEEEproof}

 We are now in position to state our main contribution which is summarized by the next proposition that introduces an adaptive PI-PBC (API-PBC) for \eqref{Phmodel}.

\begin{proposition}
	[\textit{Adaptive PI-PBC}] Consider the closed loop of the FC System modeled by eqs. \eqref{Phmodel}, the parameter estimator \eqref{gd_fc}-\eqref{iandi_fc} and the API-PBC
	\begin{subequations}\label{apipbc}
		\begin{align}
			u=&- K_p\hat y_N - K_Ix_c,\\
			\dot x_c =& \hat y_N,
		\end{align}
	\end{subequations}
	where $\hat y_N (x) := \hat x_2^\star x_3  - x_3^\star x_2$, $K_p$ and $K_I$ are positive tunning gains,  $x_3^\star>0$ is the voltage setpoint and $\hat x_2^\star$  is the estimation of $x_2^\star$ computed from \eqref{eseq}. Suppose  that \eqref{C_2} together with Assumptions \ref{ass1} and \ref{ass2} are satisfied. Then, all signals remain bounded and  the equilibrium point $(\hat{\theta}, x,x_c)=(\theta, x^\star,-K_I^{-1}u^\star)$ is asymptotically stable.
\end{proposition}

\begin{IEEEproof}
	Note that the adaptive PI-PBC \eqref{apipbc} depends on the estimate $\hat{x}^\star_2$. It can be represented by the function $u=\beta(x,\hat{x}_2^\star,x_3^\star,x_c)$. We write $\hat{x}_2^\star$ as the addition $\hat x^\star_2=x^\star_2+\delta$, where $\delta:=\hat{x}^\star_2 - x^\star_2$ is the deviation of the estimation of the equilibrium with respect to its actual value. The dynamics of the FC System in a closed loop with the controller has the form
	\begin{equation}\label{s}
		\begin{aligned}
			\dot{x}=& f_{cl}( x, \beta (x,x^\star_2+\delta,x_3^\star,x_c)),\\
			\dot{x}_c=&g_{cl}(x,x^\star_2+\delta,x_3^\star).
		\end{aligned}
	\end{equation}
	Setting $\delta =0$ in \eqref{s} results in the system of Lemma \ref{pi} which has been proven to be exponentially stable. On other hand, from Proposition \ref{prop1}, $\hat{\theta}$ is bounded and converges to $\theta$.  From Lemma \eqref{eqexistence}, $\hat{x}_2^\star$, a root of $p(x_2,\hat{\theta})=0$, is guaranteed to exist in a neighborhood of $\theta$. The error $\delta$ is then bounded and $\delta\to 0$ as $\hat{\theta}\to\theta$. Now, in virtue of  \cite[Corollary 9.2]{terrell}, the equilibrium point  $(x,x_c)=(x^\star,-K_I^{-1} u^\star)$ of \eqref{s} is asymptotically stable.
\end{IEEEproof}

\section{Experimental results}

\begin{figure}[t!]
    \centering
    \includegraphics[width=1.00\linewidth]{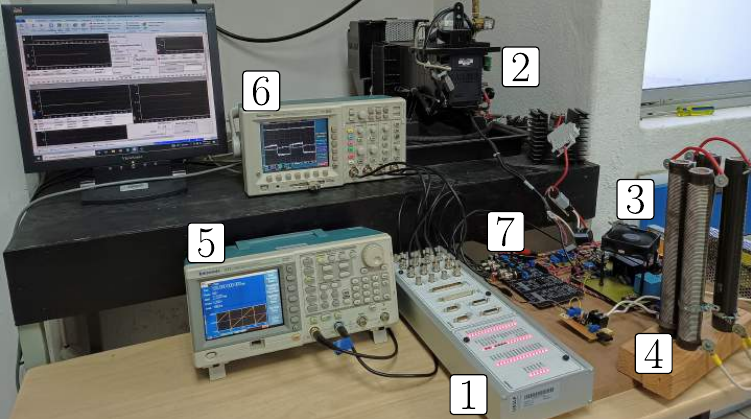}
    \caption{Experimental testbench for the FC System with the Nexa\textregistered   $ $ PEMFC power stack.}
    \label{fig:Testbench}
\end{figure}

\begin{figure*}[t!]
	\centering
	\includegraphics{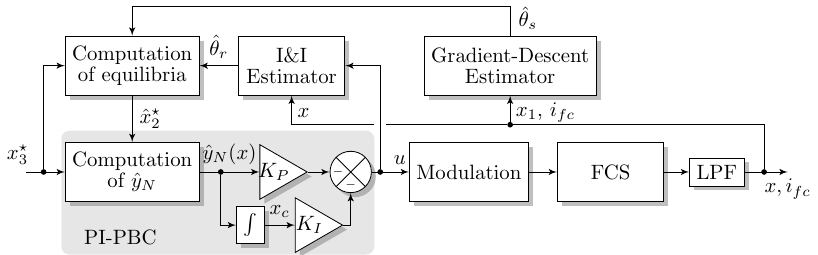}
	\caption{Implementation diagram of the API-PBC}\label{imp_diag}
\end{figure*}

Experimental validation of the API-PBC for output voltage regulation is performed with the test bench shown in Fig. \ref{fig:Testbench}. It consists of: 
1) a data acquisition system dSPACE-DS1104, 2) a fully automated 1.2 kW Nexa$\textregistered$ PEMFC power stack, 3) a 250 W boost converter prototype, 4) a resistive load, 5) a function generator, 6) an oscilloscope, and 7) a conditioning circuit boards for voltage and current sensing. The dSPACE is configured with the Euler numerical solver and a fixed-time step of 100 $\mu$s. The nominal values of the experimental setup and the gain values are shown in Table \ref{tab:parameters}. The open-circuit voltage of the PEMFC was manually measured before starting the experimentation session. It reached a value of $E_{oc}=38.84$ V. An implementation block diagram of the API-PBC is illustrated in Fig. \ref{imp_diag}. In this figure, the block labeled as “Computation of equilibria” receives the parameter estimate $\hat\theta$. From this vector, an estimation of the equilibrium point $\hat x^\star_2\in\hat{\mathcal{E}}$ is computed via the Newton-Raphson Method. That is, a numerical solution of
$$p(x_2,\hat{\theta})=0,$$
is online found, for each value $\hat{\theta}$ given by the estimation algorithm.

\begin{table}[b!]
\centering
\begin{tabular}{ccc|cc}
\hline
Parameters & Nominal value & Unit      & Gains     & Values              \\ \hline
$C$            & 136           & $\mu$F    & $k_1$     & 2.00                \\
$C_{fc}$       & 5.19          & mF        & $k_2$     & 2.00                \\
$L$            & 38.6          & $\mu$H    & $K_P$     & 19.0$\times10^{-6}$ \\
$\theta_{r1}$  & 8.30          & m$\Omega$ & $K_I$     & 0.28                \\
$\theta_{r2}$  & 47.1 or 94.2  & mS        & $\lambda$ & 4.50                \\
$f_{sw}$       & 100           & kHz       & $\gamma$  & 3.00                \\
$E_{oc}$       & 38.84         & V         &           &                     \\ \hline
\end{tabular}
\caption{Parameters and gains used during experimental validation.}
\label{tab:parameters}
\end{table}


\noindent The correct performance of the API-PBC for output voltage regulation is verified against two standard and real-life scenarios. First, pulsating changes are considered in the voltage reference while the load is kept constant. Second, pulsating changes are considered in the load while the voltage reference is kept constant. The results of both scenarios are detailed below.

\begin{figure}[t!]
	\centering
        \begin{subfigure}[Online estimations with voltage reference changes. ]
            {\includegraphics[]{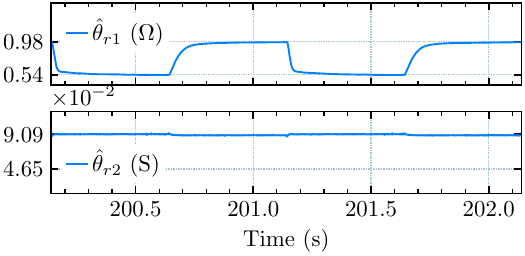}\label{fig:i&i:vref}}     
        \end{subfigure}
    \hfill
        \begin{subfigure}[Online estimations with load changes.]
		{\includegraphics[]{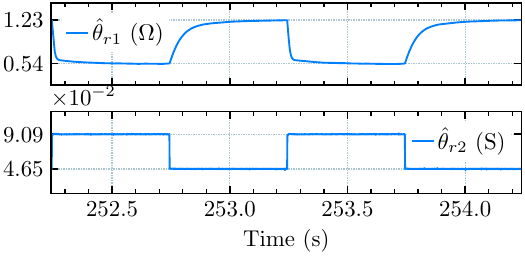}\label{fig:i&i:load}}
        \end{subfigure}
	\caption{Experimental results of the online estimation of the parasitic resistance of the inductor and the load conductance through I\&I.}\label{fig:i&i}
\end{figure}


\subsubsection{Pulsating voltage reference changes}
\begin{figure}[t!]
	\centering
        \begin{subfigure}[From top to bottom: $x_3$ (Ch1), $x_3^\star$ (Ch2), $u$ (Ch3), and $i_{fc}$ (Ch4). ]
		{\includegraphics[width=.5\linewidth]{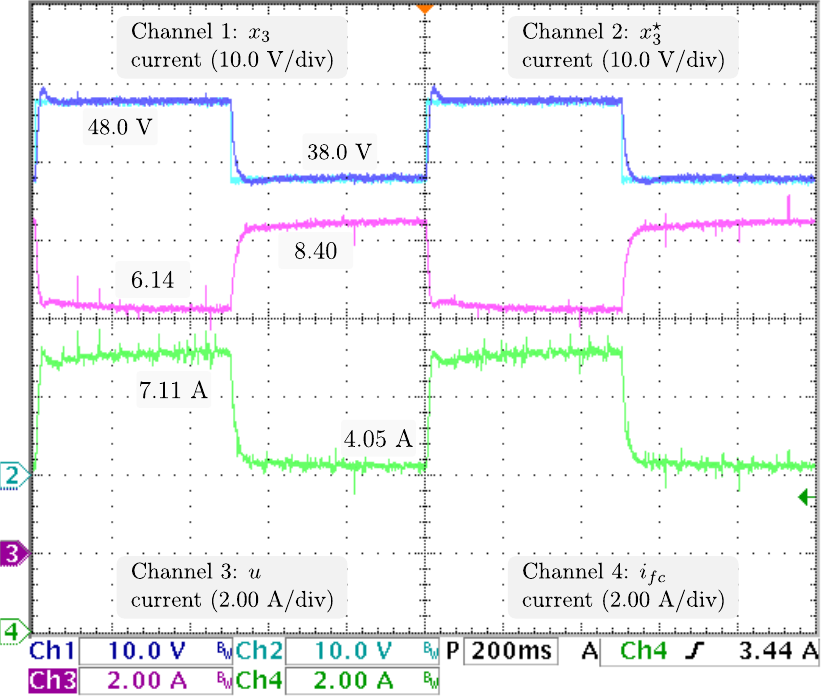}\label{fig:vref:A}}
	\end{subfigure}
    \hfill
        \begin{subfigure}[From top to bottom: $x_1$ (Ch1), $x_1^\star$ (Ch2),  $x_{2}$ (Ch3), and $x_2^\star$ (Ch4).]
		{\includegraphics[width=.5\linewidth]{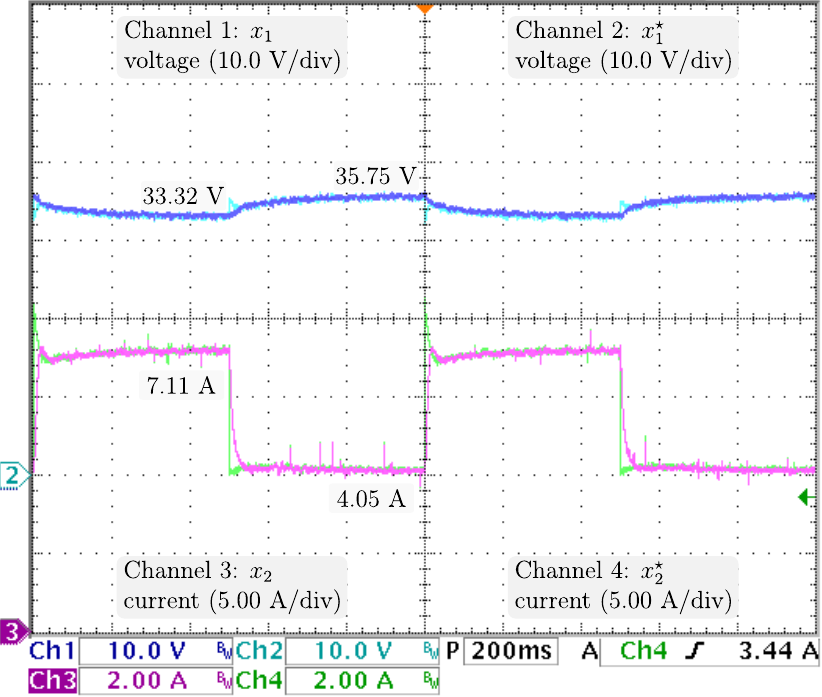}\label{fig:vref:B}}
	\end{subfigure}
	\caption{Experimental results of the output voltage regulation under pulsating voltage reference changes at 1 Hz. The resulting equilibria are $x^\star=($33.32 V, 7.11 A, 48.0 V$)$ and $x^\star=($35.75 V, 4.05 A, 38.0 V$)$. }\label{fig:vref}
\end{figure}

The voltage reference pulsates at 1.0 Hz from 48.0 V to 38.0 V while the load is kept constant at 90.15 mS.
This scenario is performed from 141.7 s to 208.7 s.
In Figs. \ref{fig:vref}, \ref{fig:i&i:vref}, and \ref{fig:exp:grad} we present the dynamic response of the system states, PEMFC current, control signal, and online estimations. 
It is important to reiterate that voltage regulation requires a suitable estimation of the current reference $\hat x_2^\star$, which requires the online estimations of \eqref{gd_fc}-\eqref{iandi_fc} and is the numerical solution of the polynomial of \eqref{eqp}.
Inspection of Figs.  \ref{fig:vref:A} and \ref{fig:vref:B} indicate that each state is tracking its reference until the voltage reference changes, causing a significant error in the output voltage.
After this, in less than 80 ms the output voltage is again tightly regulated and the other states track their references.
This implies that a suitable current $\hat x_2^\star$ is obtained through the self-tuning of the estimator and the solution of \eqref{eqp}.  
Moreover, it is also observed that the PEMFC voltage reference is being appropriately estimated with the power function model and values taken by the control signal are achievable. 
As can be seen in Fig. \ref{fig:i&i:vref}, the estimate $\hat\theta_{r1}$ is not constant, since it captures not only the parasitic resistance of the inductor, but also unmodeled resistances such as diode and switch ON resistances, and capacitor equivalent series and leakage resistances. 
On the other hand, as expected, the estimate $\hat\theta_{r2}$ has no significant variations. 
The online estimation of $\hat\theta_{s}$ can be observed in Fig. \ref{fig:exp:grad}.   
As mentioned previously, only $\hat\theta_{s2}$ is estimated online. This estimation takes a minimum of 0.680 and a maximum of 0.997 during this scenario. 
Calculating average values of the data, it is obtained $\bar{\hat \theta}_s = (0.984,\; 0.865)$. Observe that this estimation has small variations due to the hysteresis phenomenon exhibited by the PEMFC, although this fact is of no consequence for the estimation of the current reference $\hat x_2^\star$ or the PEMFC voltage $\hat x_1^\star$. 
Finally, the resulting steady-state values are 
$\hat x^\star=(33.32\;\mathrm{V}$, $7.11 \;\mathrm{A}, 48.0\;\mathrm{V})$ 
with 
$\hat \theta_r^\star=(542.31\;\mathrm{m}\Omega, 90.85\;\mathrm{mS})$
and 
$\hat x^\star=(35.75\;\mathrm{V}, 4.05\;\mathrm{A}, 38.0\;\mathrm{V})$ 
with 
$\hat \theta_r^\star=(978.28\;\mathrm{m}\Omega, 89.50\;\mathrm{mS})$.




\subsubsection{Pulsating load changes}
\begin{figure}[t!]
	\centering
        \begin{subfigure}[From top to bottom: $x_3$ (Ch1), $x_3^\star$ (Ch2), $u$ (Ch3), and $i_{fc}$ (Ch4). ]
		{\includegraphics[width=.5\linewidth]{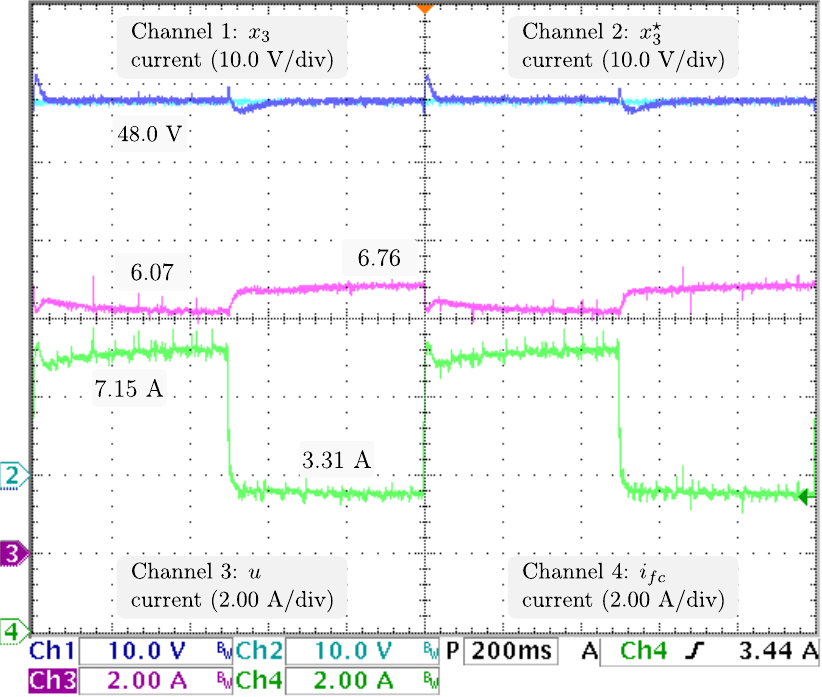}\label{fig:load:A}}
	\end{subfigure}
    \hfill
        \begin{subfigure}[From top to bottom: $x_1$ (Ch1), $x_1^\star$ (Ch2),  $x_{2}$ (Ch3), and $x_2^\star$ (Ch4).]
		{\includegraphics[width=.5\linewidth]{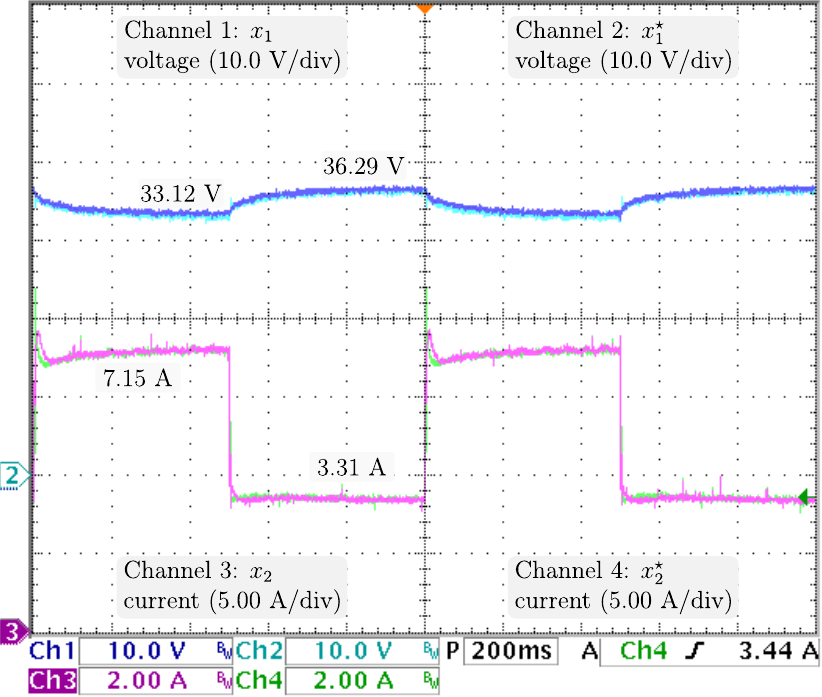}\label{fig:load:B}}
	\end{subfigure}
	\caption{Experimental results of the output voltage regulation under pulsating voltage reference changes at 1 Hz. The resulting equilibria are $x^\star=($33.12 V, 7.15 A, 48.0 V$)$ and $x^\star=($36.29 V, 3.31 A, 48.0 V$)$. }\label{fig:load}
\end{figure}

\begin{figure*}[h!]
    \centering
    \includegraphics{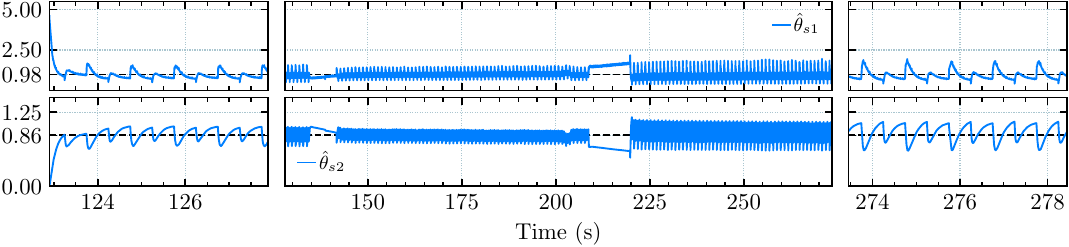}
    \caption{Experimental results of the online PEMFC parameters estimations through gradient-descend.}
    \label{fig:exp:grad}
\end{figure*}

\noindent  The load pulsates at 1.0 Hz from 90.87 mS to 46.54 mS while the voltage reference remains at 48.0 V. 
This scenario is performed from 219.9 s to 278.4 s.
In Figs. \ref{fig:load}, \ref{fig:i&i:load}, and \ref{fig:exp:grad}, we present the dynamic response of the system states, PEMFC current, control signal, and online estimations. 
As we can see from Figs. \ref{fig:load:A}, \ref{fig:load:B}, and \ref{fig:i&i:load}, each state is tracking its reference until the load changes, which produces a noticeable error in the output voltage. 
Then, in less than 120 ms, the output voltage is again tightly regulated and the other states track their references. This implies that a suitable current $\hat x_2^\star$ is obtained through the self-tuning of the hybrid estimator and the solution of \eqref{eqp}.  
Also, similar to the previous scenario, the PEMFC voltage is properly estimated and the values taken by the control signal are achievable. 
Similar to the previous scenario, we observe from Fig. \ref{fig:i&i:load} that $\hat\theta_{r1}$ has smooth variations, while the estimation $\hat\theta_{r2}$ is done quickly. 
The online estimation of $\hat\theta_{s}$ can be found in Fig. \ref{fig:exp:grad}. In this scenario, the online estimation $\hat\theta_{s2}$ takes a minimum of 0.600 and a maximum of 1.167.
Computing average data values $\bar \theta_s^\star = (0.999,\; 0.921)$. Note that in this scenario we also see variations in the estimates.
Finally, the resulting steady-state values are $\hat x^\star=(33.12\;\mathrm{V}, 7.15\;\mathrm{A}, 48.0\;\mathrm{V})$, with 
 $\hat \theta_r^\star=(536.18\;\mathrm{m}\Omega, 90.87\;\mathrm{mS})$
and 
$\hat x^\star=($36.29$\;\mathrm{V}$, 3.31$\;\mathrm{A}, 48.0\;\mathrm{V})$, with
$\hat \theta_r^\star=(1.22\;\Omega, 46.54\;\mathrm{mS})$.

\begin{figure}[t!]
    \centering
    \includegraphics{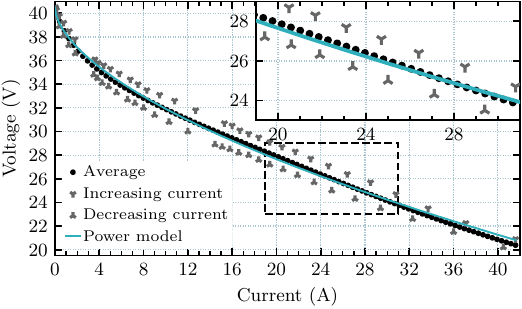}
    \caption{Comparison of experimental ``Data'' with the polarization curve model computed using the average values for the estimations in both scenarios, ``Scenario 1'' and ``Scenario 2''.}
    \label{fig:exp:polcurve}
\end{figure}

Lastly, in Fig. \ref{fig:exp:polcurve} we compare experimental ``data'' obtained of more than ten thousand $(i_{fc}, v_{fc})$ measurements with a model of the polarization curve for each scenario, ``Scenario 1'' and ``Scenario 2''. These models are computed using the measured $E_{oc}$ and their respective averaged $\bar{\hat\theta}_s$. As we can see, the hysteresis phenomena is observed in the experimental data, and is worth mentioning that modeling this complex dynamics is beyond the scope of this study. 
Nevertheless, in steady-state operation the power model ``Scenario 1'' fits  the central section of the hysteresis band, meanwhile the power model of ``Scenario 2'' fits  only at low currents. 
It should be noted that average values $\bar{\hat\theta}_s$ obtained off-line were used only for the comparison. Recall that online estimation of $\hat\theta_s$, as shown in Fig. \ref{fig:exp:grad}, is given to the API-PBC and during both scenarios a suitable estimation of  the inductor current and PEMFC voltage equilibrium points are obtained.

\section{Conclusions}
We derived a PI-PBC for the fuel-cell system of eq. \eqref{Phmodel}. 
The resulting controller has the same structure as that in  \cite{pimike} in spite of the fact that the class of systems to which the FC System belongs is not the one therein addressed.
An adaptive version of the approach is afterward introduced, based on an indirect output voltage regulation control scheme.
Parameter estimation is carried out using the I\&I and the gradient-descent online algorithms. 


As a final comment, we notice that system \eqref{Phmodel} can be generalized to systems having the form
\begin{align}\label{gen}
 Q\dot x = [J_0-R]x + \sum_{i=1}^{m}g_i(x)u_i + d(x),
\end{align}
where $d:\mathbb{R}^n\to\mathbb{R}^n$ and
$$g_i(x)=J_ix+b_i.$$
Clearly, \eqref{gen} enlarges the family of systems considered in \cite{pimike}. On the other hand, we identify that the two key features that make possible  to apply the PI-PBC  presented in \cite{pidzonetti,pimike} to the system \eqref{Phmodel} are the monotonicity property \eqref{mon} and the equality \eqref{cond}.

In this manner, an immediate attempt to extend Lemma \ref{pi} would lead to the imposition on system \eqref{gen} of the following two (restrictive) conditions:

\begin{enumerate}
    \item[C1.] The $\ell$-th entry of vector $d$ is the mapping $d_\ell(x_\ell)$, which satisfies strong monotonicity. That is, for any scalars $y$ and $z$ there exists a non-negative constant $\alpha_\ell$  such that $$(y-z)^\top \big[[-d_\ell(y)] - [-d_\ell(z)] \big ] \geq \alpha_\ell |y-z|^2,$$ with $\ell=1,\cdots ,n$.
    \item[C2.]  For $y$ and $z\in\mathbb{R}^n$ and $i=1,\cdots,m$, $$g_i^\top (y)Q^{-1}d(z)=0.$$
\end{enumerate}

 Future work is oriented to the relaxation of conditions C1 and C2 in order for the PI-PBC to be applied.

\bibliography{bibliography}
\bibliographystyle{IEEEtran}

\end{document}